\documentclass[useAMS,usenatbib]{mn2e}
\usepackage{graphicx}
\usepackage{amssymb}
\usepackage{amsmath}
\usepackage{natbib}
\usepackage[section]{placeins}
\usepackage{verbatim}

\usepackage{booktabs}

\usepackage{mathabx}
\usepackage{url}

\def\Rsol{R$_\odot$}

\newcommand{\bjdtdb}{\ensuremath{\rm {BJD_{TDB}}}}

\newcommand{\rprs}{$R_{\rm{pl}}/R_{\star}$}

\makeatletter

\newcommand{\Rmnum}[1]{\expandafter\@slowromancap\romannumeral #1@}
\makeatother

\title[Detection of potassium in HAT-P-1b from narrowband spectrophotometry]{GTC OSIRIS transiting exoplanet atmospheric survey: detection of potassium in HAT-P-1b from narrowband spectrophotometry\thanks{Based on observations made with the Gran Telescopio Canarias (GTC), installed in the Spanish Observatorio del Roque de los Muchachos of the Instituto de Astrof\'isica de Canarias, in the island of La Palma, and part of the large ESO program 182.C-2018.}}

\author[P. A.~Wilson et al.]
{\parbox{\textwidth}{P. A. Wilson$^{1,2,3}$\thanks{E-mail: \texttt{pwilson@iap.fr}},
D.~K.~Sing$^{3}$,
N.~Nikolov$^{3}$,
A.~Lecavelier des Etangs$^{1,2}$,
F.~Pont$^{3}$,
J.~J.~Fortney,$^{4}$
G.~E.~Ballester$^{5}$,
M.~L{\'o}pez-Morales$^{6}$,
J.-M.~D{\'e}sert$^{7}$,
A.~Vidal-Madjar$^{1,2}$}\vspace{0.4cm}\\
\parbox{\textwidth}{
$^{1}$CNRS, UMR 7095, Institut d'Astrophysique de Paris, F-75014, Paris, France\\
$^{2}$Sorbonne Universit{\'e}s, UPMC Univ Paris 06, UMR 7095, Institut d'Astrophysique de Paris, F-75014, Paris, France\\
$^{3}$Astrophysics Group, School of Physics, University of Exeter, Stocker Road, Exeter EX4 4QL, UK\\
$^{4}$Department of Astronomy and Astrophysics, University of California, Santa Cruz, CA 95064, USA\\
$^{5}$Lunar and Planetary Laboratory, University of Arizona, Sonett Space Sciences Building, Tucson, Arizona 85721-0063, USA\\
$^{6}$Harvard-Smithsonian Center for Astrophysics, 60 Garden Street, Cambridge, Massachusetts
02138, USA\\
$^{7}$Department of Astrophysical and Planetary Sciences, University of Colorado, Boulder CO 80309}}

\begin{document}

\date{Accepted 2015 March 23. Received 2015 March 17; in original form 2015 February 17}

\pagerange{\pageref{firstpage}--\pageref{lastpage}} \pubyear{2002}

\maketitle

\label{firstpage}

\begin{abstract}
We present the detection of potassium in the atmosphere of HAT-P-1b using optical transit narrowband photometry. The results are obtained using the 10.4-m Gran Telescopio Canarias ({\it GTC}) together with the OSIRIS instrument in tunable filter imaging mode. We observed four transits, two at continuum wavelengths outside the potassium feature, at $6792$~\AA\ and $8844$~\AA, and two probing the potassium feature in the line wing at $7582.0$~\AA\ and the line core at $7664.9$~\AA\ using a 12~\AA\ filter width ($\mathrm{R}\sim650$). The planet-to-star radius ratios in the continuum are found to be \rprs$ = 0.1176 \pm 0.0013$ at $6792$~\AA\ and \rprs$ = 0.1168 \pm 0.0022$ at 8844~\AA, significantly lower than the two observations in the potassium line: \rprs$= 0.1248 \pm 0.0014$ in the line wing at $7582.0$~\AA\ and \rprs$ = 0.1268 \pm 0.0012$ in the line core at $7664.9$~\AA. With a weighted mean of the observations outside the potassium feature \rprs$= 0.1174 \pm 0.0010$, the potassium is detected as an increase in the radius ratio of $\Delta$\rprs$ = 0.0073\pm0.0017$ at $7582.0$~\AA\  and $\Delta$\rprs$ = 0.0094\pm0.0016$ at $7664.9$~\AA\ (a significance of 4.3 and 6.1~$\sigma$ respectively). We hypothesize that the strong detection of potassium is caused by a large scale height, which can be explained by a high-temperature at the base of the upper atmosphere. A lower mean molecular mass caused by the dissociation of molecular hydrogen into atomic hydrogen by the EUV flux from the host star may also partly explain the amplitude of our detection.
\end{abstract}


\section{Introduction}
Transiting exoplanets provide astronomers with a unique opportunity to study the dynamics, structure and composition of exoplanetary atmospheres. A valuable insight into the planet's atmosphere is gained by measuring the stellar light transmitted through the exoplanet atmosphere as a function of wavelength. The amount of light obscured by the transiting exoplanet is dependent on the composition of its atmosphere, with absorbing or scattering species letting less light through at specific wavelengths, causing a change in the exoplanet radius as a function of wavelength. Hot Jupiters in particular, a subgroup of exoplanets that orbit close to their host star and exhibit inflated radii have proven invaluable in the characterisation of exoplanet atmospheres.

Hot Jupiters have thus far shown great atmospheric diversity with some planets being obscured by high altitude hazes or clouds (e.g. HD~189733b, \citealt{pont13} and WASP-12b, \citealt{sing13}) whilst others have shown relatively clear atmospheres (e.g. HD~209458b, \citealt{sing08}, WASP-19b, \citealt{huitson13} and HAT-P-11b, \citealt{fraine14}) with clear signs of alkali metals at optical wavelengths and water features at near-IR wavelengths. Amongst the sub-group of hot Jupiters that are not dominated by clouds and hazes, the most dominant sources of opacity in the optical are predicted to be the alkali metals sodium and potassium \citep{seager00}, with their strong resonance doublets at 5890, 5896~\AA\ and 7665, 7699~\AA\, respectively, with Rayleigh scattering by H$_2$ dominating at shorter wavelengths. Sodium was first detected by \cite{charbonneau02} in the atmosphere of HD~209458 b using the Space Telescope Imaging Spectrograph (STIS) on the Hubble Space Telescope ({\it HST}). These observations were later confirmed by \cite{snellen08} using the High Dispersion Spectrograph at the Subaru telescope, with the sodium line profile characterized by \cite{sing08} using the {\it HST}. Exoplanetary sodium has also been found to be present in the atmospheres of HD~189733b \citep{redfield08,jensen11,huitson12}, WASP-17b \citep{wood11,zhou12}, XO-2b \citep{sing12} and HAT-P-1b \citep{nikolov14}. Exoplanetary potassium has been detected on the hot Jupiters XO-2b \citep{sing11} and WASP-31b \citep{sing15}.

Tunable Filters (TFs) have the unique capability of having the central wavelength and filter passband tuned to a specific value. TFs consist of a Fabry-P\'erot etalon made up of two parallel reflective surfaces. As a result of carefully varying the separation between the two plates, the filter width and central wavelength can be accurately selected. TF have several advantages over low resolution spectroscopy. They provide accurate differential photometry and have the unique advantage that they can be tuned to wavelengths not contaminated by strong telluric lines such as the prominent O$_2$ lines near 6884 and 7621~\AA\ \citep{catanzaro97}. Since no diffraction gratings are used, TFs can also be much more efficient \citep{colon10}, especially for observing atomic absorption features. Combining this technique with the 10.4~m aperture of the {\it GTC} telescope makes it possible to study the atmospheres of planets orbiting stars such as HAT-P-1 ($\mathrm{V}=9.87$) which are much fainter than HD~209458 ($\mathrm{V}=7.63$) and HD~189733 ($\mathrm{V}=7.65$), which due to their apparent brightness, and large scale heights, are the two best studied cases thus far.

In this study we present the detection of potassium in the atmosphere of HAT-P-1b \citep{bakos07}, a $1.319~R_{\rm{Jup}}$ exoplanet with a mass of $0.525~M_{\rm{Jup}}$, and thus a low average density of $\rho = 0.345$~g/cm$^3$, on a 4.47~day circular orbit around a G0V star at a distance of 0.055~au \citep{bakos07,johnson08}. HAT-P-1B (\mbox{BD+37 4734B}) is a part of a visual binary system with the F8 companion star \mbox{BD+37 4734A}, located more than 450 light years away in the constellation of Lacerta. The host star appears to not be very active (see \S~\ref{sec:stellar_activity}) \citep{knutson10,nikolov14}. The planet HAT-P-1b shows signs of a modest temperature inversion layer \citep{todorov10}. {\it HST}/STIS observations of HAT-P-1b by \cite{nikolov14} detected the sodium doublet, yet found no sign of potassium nor the pressure-broadened wings of the two alkali metals. Instead they found a flat optical spectrum, and a tentative absorption enhancement at wavelengths longer than $\sim0.85\mu$m. The {\it HST} Wide Field Camera 3 (WFC3) observations by \cite{wakeford13} detected a significant water absorption feature in the 1.4~$\mu$m absorption band. The results by \cite{nikolov14} indicate a strong optical absorber at higher altitudes.

With its low density and large radius, HAT-P-1b is an interesting test case for interior and atmosphere models. These observations are a part of our larger spectrophotometric survey aimed at detecting and comparing atmospheric features in transiting hot Jupiters (ESO programme 182.C-2018). In \S~2 we describe the observations, and in \S~3 we describe the analysis of the transit light curves. In \S~4 we present a discussion of the results where we compare the observations to previous {\it HST} observations, and conclude in \S~5.

\section{Observations}
\label{obs}
Observations were performed using the 10.4~m {\it GTC} telescope located at Observatorio del Roque de los Muchachos of the Instituto de Astrof\'{i}sica de Canarias on the island of La Palma. Narrowband imaging was done with the OSIRIS instrument using the red tuneable filter (operating range of 6510~\AA\ - 9345~\AA) tuned to the minimum width of 12~\AA.

\subsubsection*{Instrumental setup}
The OSIRIS instrument \citep{cepa03,cepa00,cepa98} consists of a mosaic of two Marconi CCD42-82 CCDs each with a 2048 x 4096 pixel detector separated by a 72 pixel gap between them. Each pixel has a physical size of 15~$\mu$m, which gives a plate scale of $0\farcs125$. The observations were performed without any binning with a readout frequency of $500$ kHz and a gain of $1.46$ e$^-$/ ADU on CCD2. This setup gives a readout noise of $8$e$^{-}$. The observations tuned to $6792$~\AA\ were performed using the whole CCD2 array. Being one of the first OSIRIS observations, a sub-array mode was not offered. Later observations have since been performed in the sub-array mode as it reduces the read time yielding a higher cadence, and because the photometry of the first observations was not improved by including other fainter stars in the HAT-P-1 field. The observations tuned to 7582.0 and 7664.9~\AA\ were windowed to 600x700 pixels. To obtain the highest possible resolution the smallest possible passband of 12~\AA\ was chosen.

\subsubsection*{Observing log}
HAT-P-1b was observed on four separate nights. For all observations, the companion star HAT-P-1-B (\mbox{BD+37 4734A}) was used for photometric comparison due to its close $\sim11$\arcsec proximity allowing for windowed frames to be taken. The stars are of similar spectral type and brightness. The frames were also rotated to ensure that both the target and the comparison star were at the same radial distance from the centre of the CCD, such that both objects were observed at the same wavelength. Due to problems with the {\it GTC} dome shutter, targets with an elevation above 72 degrees are subject to vignetting. Since the $22$ October 2009 or the $19$ July 2011 observations crossed this 72 degree limit, reaching the maximum elevation of HAT-P-1b at 80 degrees, a decrease in the raw target flux was expected. Despite this, no signs of vignetting were observed.

{\it $22$ October 2009:}
The tunable filter was centred on the continuum at $6792$~\AA. The observations began at 20:42 UT, and ended at 02:25 UT. Due to variable seeing, ranging from about 1.4 to 0.7 arc seconds, the exposure time was frequently adjusted to avoid saturation from 10 to 20s. Being the first observations with the OSIRIS instrument in our program, there were still issues present concerning the dark current. A reduction of the data was done both with and without dark frames. Including the darks increased the photometric noise slightly and were thus not used in the reduction. We conclude that reducing the data using dark frames only introduces noise for such short exposure times. Hardware upgrades have since solved the issue.

{\it $19$ November 2010:}
The tunable filter was  centred at $7582.0$~\AA\ and $7664.9$ \AA, resulting in near simultaneous light curves at two wavelengths sampling the line wing of the potassium feature and the core of the KI D2-line. Tuning between the two wavelengths took $0.1 \mathrm{s}$ with the cadence of the wavelength alteration set by the readout time which was $5 \mathrm{s}$. The observations started late at 22:15 UT due to high humidity and ended at 01:29 UT. Seeing was variable ranging from 0.8 to 1.3 arc seconds. The exposure times were adjusted to avoid saturation. An hour into the observations light cirrus clouds were present.

{\it $19$ July 2011:}
The tunable filter was centred on the continuum at $8844$~\AA. The observations began at 00:00 UT and ended at 04:06 UT on the 19$^{\mathrm{th}}$ of July. Seeing was very variable during the sequence, with values ranging from 0.8 to 2.2 arc seconds. After 02:40 UT, problems with the primary mirror caused a distortions of the PSF (Point Spread Function). The problem was subsequently corrected at 03:10 UT.

{\it $26$ November 2013:}
These observations were a repeat of the $19$ November 2010 observations and used the same setup. To decrease the overheads the exposure times were increased from 6 to 15 seconds. Observations started late at 22:16 (due to high humidity) and ended at 01:29 UT (lower elevation limit of the GTC of 25 degrees).

\subsection{Reductions}
The image reductions were made using standard IRAF\footnote{IRAF is distributed by the National Optical Astronomy Observatories, which are operated by the Association of Universities for Research in Astronomy, Inc., under cooperative agreement with the National Science Foundation.} routines that included bias subtraction, flat field division and the removal of cosmic rays. An average of 0.5 to 1.7 cosmic rays were detected per image, except for the 6792~\AA\ observations where an average count of 58 was detected. These events are likely not cosmic rays, but instead due to issues with the dark current. For the cosmic ray removal the IRAF routine {\tt{cosmicrays}} was used with a threshold of 100 and a flux ratio of 4. A large number of calibration frames were used which included at least 50 bias frames and 100 dome flats. Images with peak ADU counts above 45 000 were removed to ensure the photometry was performed in the linear regime of the detector.

\begin{figure*}
\centering
\includegraphics[width=60mm]{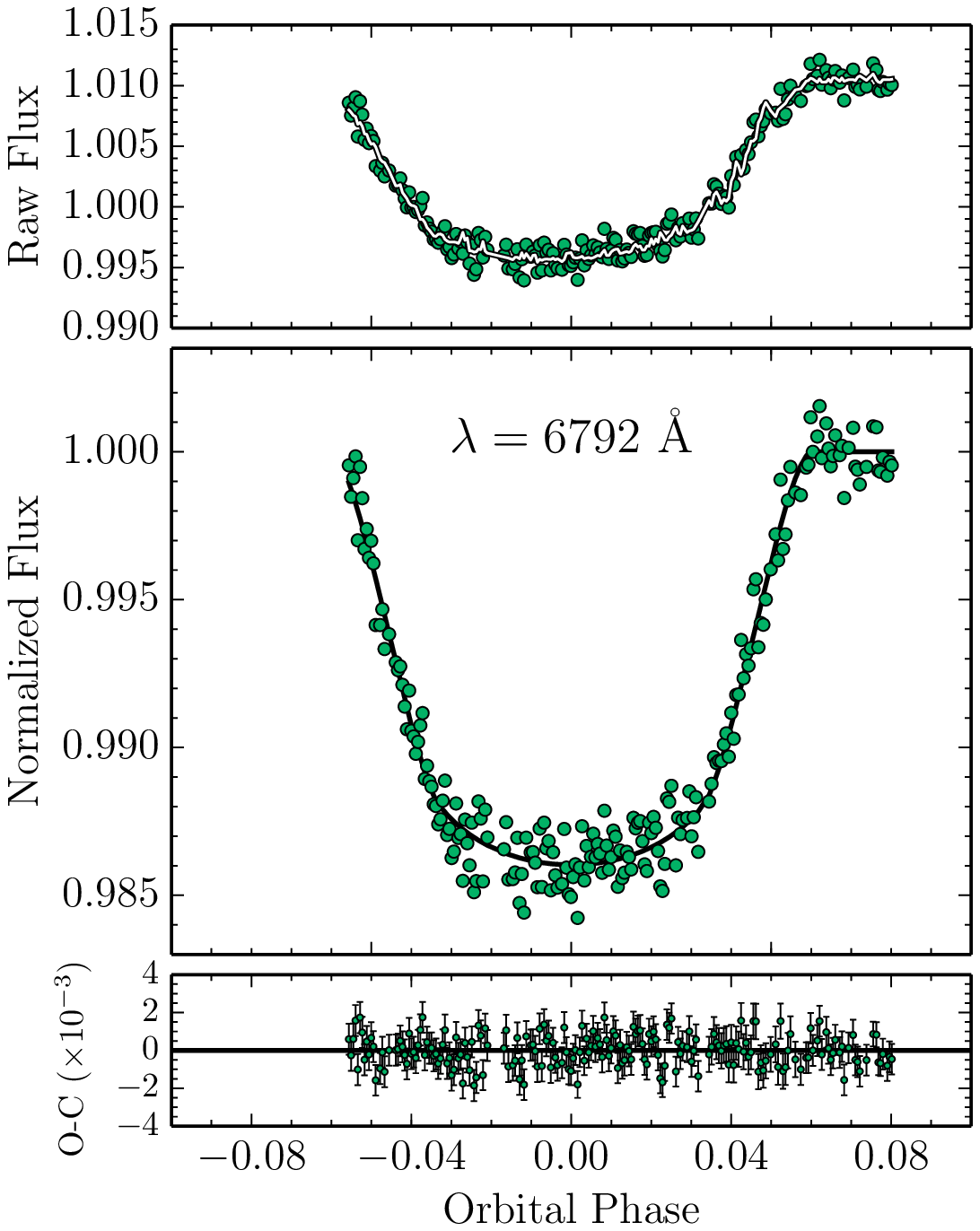}
\includegraphics[width=53.6mm]{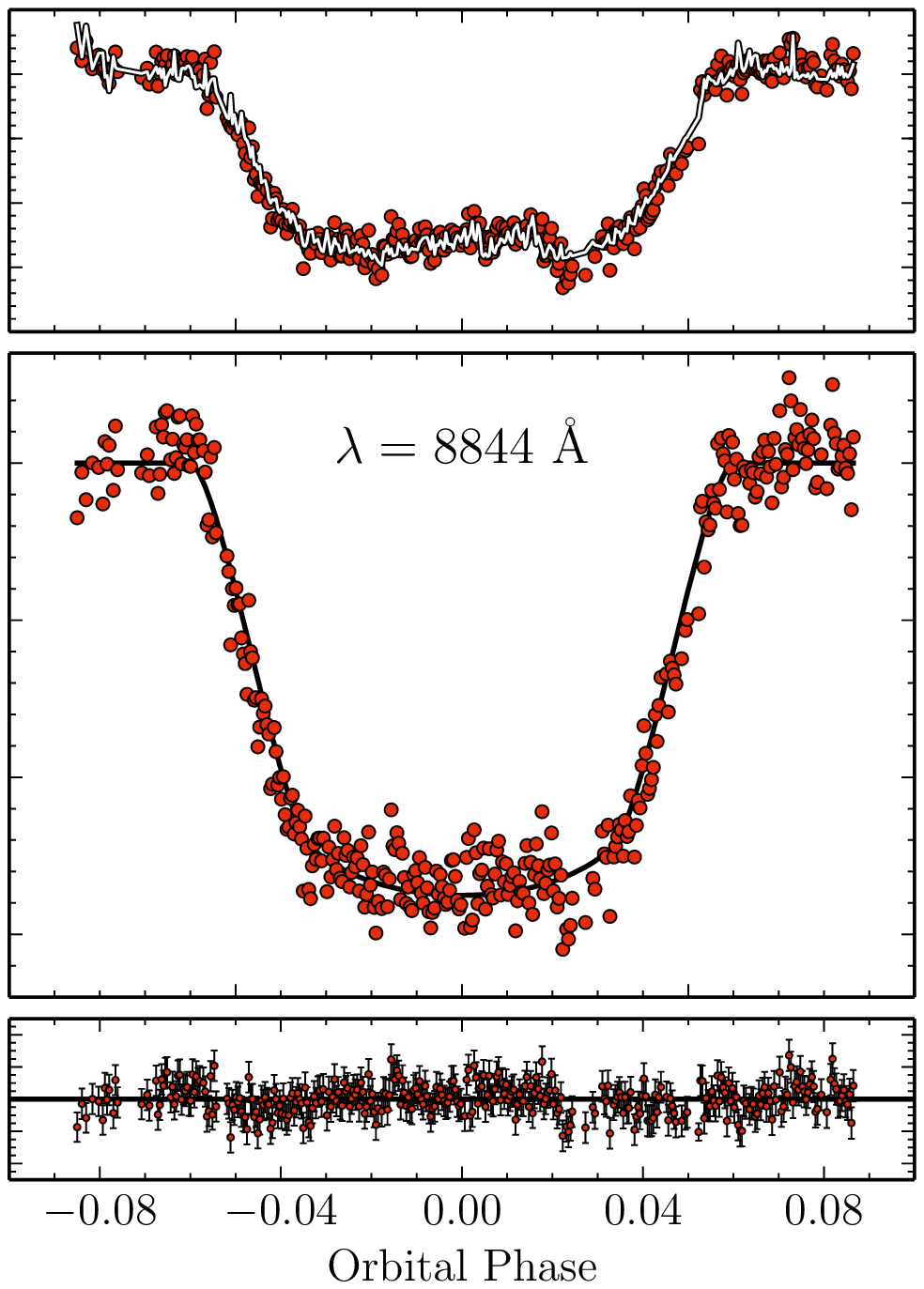}

 \caption[The 6792~\AA\ and 8844~\AA\ HAT-P-1b light curves.]{The 6792~\AA\ and 8844~\AA\ observations (outside the potassium feature) with the raw light curve shown in the top panel, the modeled light curve in the middle panel and the best fit residuals shown in the bottom panel. The systematic trends seen in the 8844~\AA\ light curve were well modeled with a linear FWHM fit which incorporated the changing PSF.}
\label{img:lcs_6797}
\end{figure*}

\begin{figure*}
\includegraphics[width=150mm]{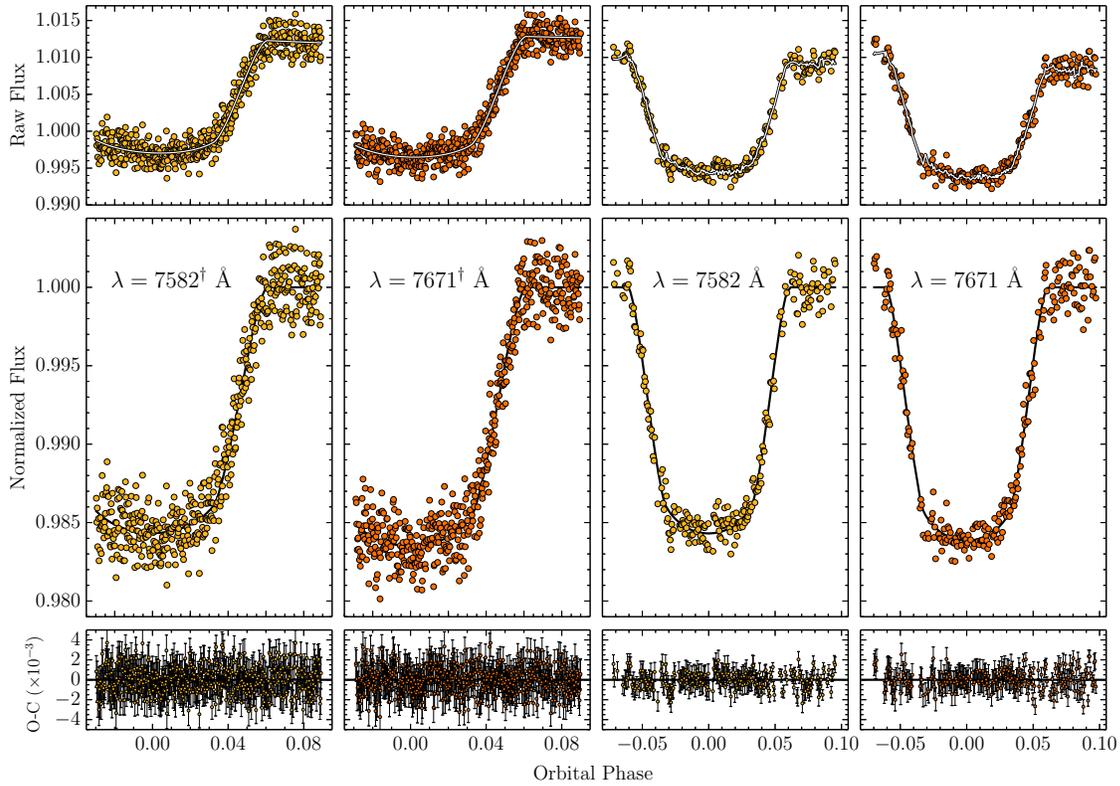}
 \caption[Light curves probing the potassium feature of HAT-P-1b.]{Light curves probing the potassium feature. The first and third columns (yellow points) corresponding to observations at 7582.0~\AA, and the second and fourth column (orange points) corresponding to the core of the potassium line at 7664.9~\AA. The half transits were obtained 19 November 2010 whilst the full transits were obtained on 26 November 2013.}
\label{img:lcs}
 \end{figure*}

Aperture photometry was done using the {\texttt{apphot}} package in IRAF using a Gaussian centring algorithm. In order to ensure the best possible photometry, a large range of apertures were explored varying both the aperture size as well as the dimensions of the sky annulus until the RMS scatter of the light curve residuals were minimised. The residuals were calculated by subtracting the data points from an optimal light curve fit generated using the Levenberg-Marquardt least-square algorithm implemented in the SciPy\footnote{\url{http://scipy.org/}} software distribution (\texttt{optimize.leastsq}) and by iteratively rejecting any data points with standard deviation $>3~\sigma$. The FWHM (Full Width at Half Maximum) values were measured using the \texttt{psfmeasure} task in IRAF. The Barycentric Dynamical Time (\bjdtdb) was obtained by computing an average of the open and close times in UTC recorded in the image headers and converted using the online tool\footnote{\url{http://astroutils.astronomy.ohio-state.edu/time/utc2bjd.html}} based upon \cite{eastman10}. The differential photometry was performed with the target and the nearby reference star at the same wavelength. The resulting light curves are shown in Fig.~\ref{img:lcs_6797} and Fig.~\ref{img:lcs}.

\section{Analysis}

\subsection{Transit light curve fits}
The transit light curves were generated using the analytical transit equations of \cite{mandel02}. Only one reference star was used as both stars have to be the same distance from the optical centre to be observed at the same wavelength. The best fitting parameters together with their associated uncertainties were determined by a Markov chain Monte Carlo algorithm (MCMC) following \cite{wilson14}. The initial starting parameters were from \citet{nikolov14}. Possible time dependent correlations with airmass, FWHM and detector position were explored. The number of correlation parameters were determined using the Bayesian Information Criterion (BIC) \citep{schwarz78} defined as:

\begin{equation}
\mathrm{BIC} = \chi^2 +k \ln n,
\end{equation}

\noindent where $k$ is the number of free parameters and $n$ is the number of data points. The correlation parameters that minimised the BIC were used to de-trend the light curves. This was for all light curves, except the half transit light curves at 7582~\AA\ and 7664.9~\AA, where we fitted a linear correlation with FWHM. For the half transit light curves no correlation parameters where used as it only increased the BIC.

\begin{figure}
\centering
\includegraphics[width=90mm]{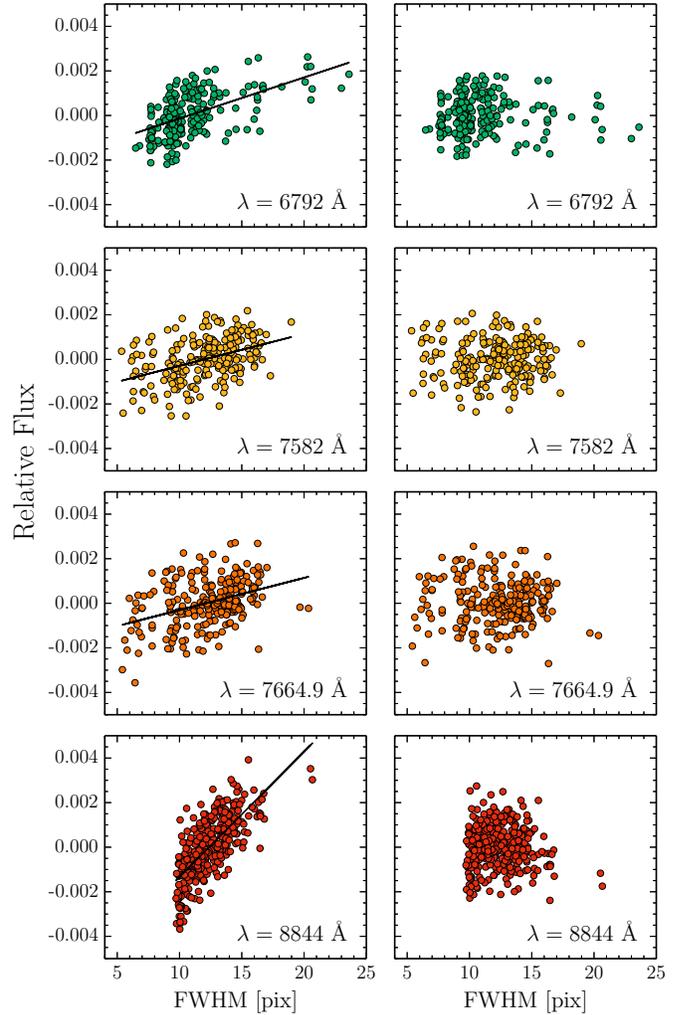}
 \caption[The residual scatter of HAT-P-1b light curves as a function of FWHM.]{The residual scatter of the HAT-P-1b light curves as a function of FWHM. The subplots on the left show a linear fit to the observed FWHM trend whilst the subplots on the right show the detrended residuals.}
  \label{img:residuals}
\end{figure}

For the observations at 6792~\AA\ we found $\Delta\mathrm{BIC}=79$, for 7582~\AA\ $\Delta\mathrm{BIC}=54$, for 7671~\AA\ $\Delta\mathrm{BIC}=60$ and for 8844~\AA\ $\Delta\mathrm{BIC}=387$ when fitting a linear correlation with FWHM using one coefficient. Additional FWHM coefficients with higher order polynomials did not improve the fit significantly with the BIC value increasing slightly and were thus rejected. The residual scatter of HAT-P-1b transit light curves as a function of FWHM is shown in Fig.~\ref{img:residuals} with the best linear fitting line shown in black.

All the transits had the planet-to-star radius ratio, $R_p/R_s$, central transit time $T_C$, a baseline normalising factor, $N$ and a slope term $s$ as free parameters. All the transits apart from the half transits had an additional free FWHM parameter. The fixed parameters taken from \citet{nikolov14} included the period, $P=4.46529976\pm(55)$~days, impact parameter $b=0.7501$ and the quadratic limb-darkening coefficients, $u_1$ and $u_2$, which were calculated using the ATLAS stellar atmospheric models\footnote{\url{http://kurucz.harvard.edu/grids.html}} following \cite{sing10}. A quadratic limb darkening law of the following form was used

\begin{equation}
\frac{I(\mu)}{I(1)} = 1-u_1(1-\mu)-u_2(1-\mu)^2,
\end{equation}
where $I(1)$ is the intensity at the centre of the stellar disk, $\mu = \cos(\theta)$ is the angle between the line of sight and the emergent intensity while $u_1$ and $u_2$ are the limb darkening coefficients. The stellar and orbital parameters were also kept fixed with $R_s= 1.174~$\Rsol, the eccentricity $e=0$ and the scaled semi-major axis $a/R_s=9.853$, the same as in \citet{nikolov14}.

Due to the different path lengths the light has to travel through the tunable filter, the observed wavelength decreases radially outward from the center of the field of view (FOV). At the wavelengths where sky emissions from the O\Rmnum{1} and OH emission lines exist, sky rings are seen. The position of these sky rings can be used to measure wavelength drifts, which may occur as the filter is being tuned between two different wavelengths. As no sky-rings were seen in the data, observing a wavelength shift becomes impossible. The typical wavelengths shifts measured in a similar study by \cite{wilson14}, who used the same setup, was found to be $\sim6$~\AA. Compared to the filter bandwidth of 12~\AA\ we conclude that any wavelength shift which may have occurred is unlikely to have a measurable impact on the derived radius ratios. 

\begin{figure}
\centering
\includegraphics[width=90mm]{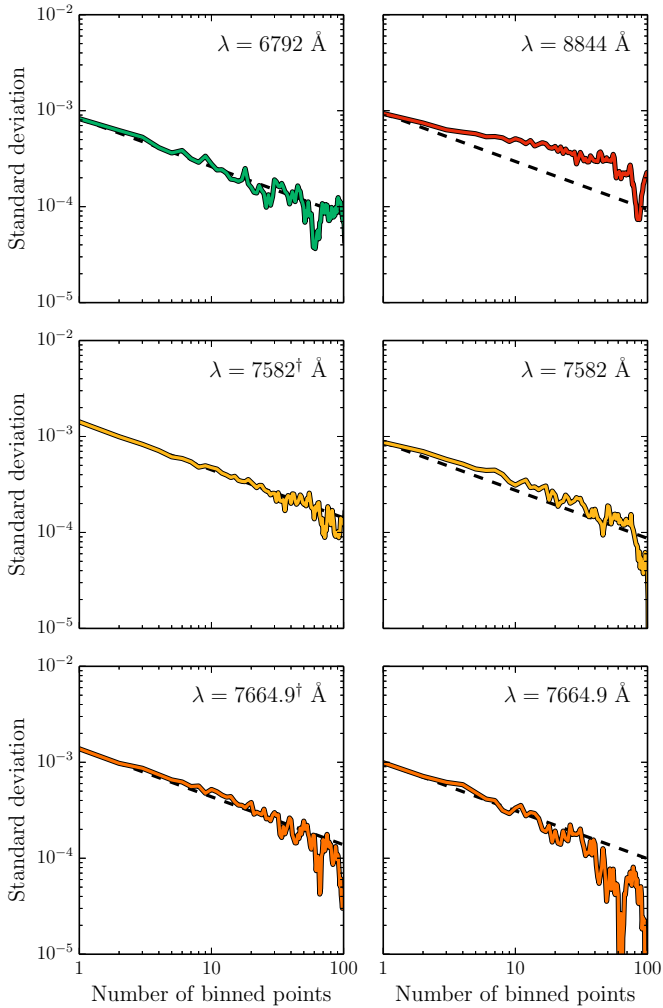}
 \caption[The residual noise scatter of the HAT-P-1b light curves.]{The residual noise scatter of the HAT-P-1b light curves (solid coloured lines), compared to the Gaussian noise expectation (dashed line). Due to large uncertainties in determining $\sigma_r$ for a large number of binned points, the red noise term was calculated as an average of red noise models using a range of $N$ from 10 to 20.}
  \label{img:red_noise}
\end{figure}

\subsection{Noise Estimate}
\label{noise}
To estimate the presence of systematic noise (red noise, $\sigma_r$) in the data, we applied the time-averaging procedure of \cite{pont06}. In the presence of uncorrelated noise (white noise, $\sigma_w$), the standard deviation of binned residuals ($\sigma_N$) are expected to follow the relation

\begin{equation}
\sigma_N = \frac{\sigma_1}{\sqrt{N}}\sqrt{\frac{M}{M-1}}
\label{sig_n}
\end{equation}

\noindent where $\sigma_1$ is the standard deviation of all the residuals, $N$ is the number of successive residuals points, $M$ the number of bins. For data with red noise, the standard deviation is expected to increase by an amount $\sigma_r$ such that for every bin containing $N$ points,

\begin{equation}
\sigma_r = \sqrt{\sigma_N^2-\frac{\sigma_w^2}{N}}.
\label{eq:sigma_r}
\end{equation}

\noindent The red noise contribution was estimated by choosing $N$ to be equal to the number of points in the transit ($n_T$), which varied in accordance with the cadence of the observations. However, due to the small number of out of transit points compared to the number of data points within the transit itself, the functional form of $\sigma_r$ as a function of the number of $N$ was calculated by averaging 10 separate fits to Eq.~\ref{eq:sigma_r} using values of $N \in [10,20]$. For $N \gtrsim 20$, the standard deviation starts to fluctuate as the number of binned points become large (see Fig.~\ref{img:red_noise}).

To take into account the effect of red noise on the radius ratio, we rescale the photometric uncertainties of the data set by a re-weighting factor, $\beta$ expressed as,

\begin{equation}
\label{eq:beta}
\beta = \sqrt{1+\left( \frac{\sigma_r}{\sigma_w} \right )^2}.
\end{equation}

\noindent following the procedures of \cite{winn09}.

We compare the estimated $\beta$ value to the values obtained using the wavelet method described in \cite{carter09} and find good agreement with previous calculated values, although the wavelet method tends to have slightly smaller beta values. We adopt a conservative approach and select the $\beta$ factor calculated using the time-averaging procedure explained above. Using the measured amount of flux within the photometry aperture, we estimate that amongst the various sources of noise, $\sim49$~--~$69$ per cent of it is due to photon noise. Co-added photometric precisions of 170~ppm were attained which is sufficiently accurate to detect atmospheric features.

\section{Results and Discussion}

\begin{figure}
\centering
\includegraphics[width=90mm]{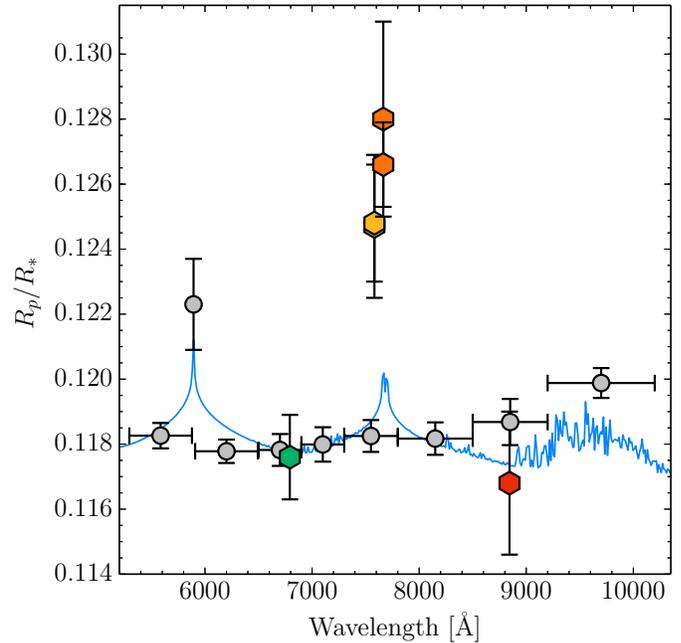}
 \caption[Transmission spectrum of HAT-P-1b (zoom in).]{Transmission spectrum of HAT-P-1b showing the STIS G750L data of \citep{nikolov14} as grey points and all the {\it GTC} data has hexagonal points. The model transmission spectrum by \cite{fortney08,fortney10} (shown as a blue line) assumes an isothermal hydrostatic uniform abundance with a equilibrium temperature of 1200~K. The model has been binned into 12~\AA\ bins and shifted vertically to best fit the {\it HST} and {\it GTC} datasets.}
  \label{img:hat-p-1b_transpec_zoom}
\end{figure}

\begin{table*}
\centering
\caption{Light curve system parameters for HAT-P-1b.}
\label{results}
 \begin{tabular}{@{}cccccccccc}
  \hline
  \hline
  Parameter & $6792$ \AA & $7582.0^{\dagger}$ \AA & $7582.0$ \AA  & $7664.9^{\dagger}$~\AA & $7664.9$ \AA & $8844$ \AA\\
  \hline
  $T_0$ [\bjdtdb] 	&  $2455127.515(20) $ & $2455520.456(57)$ & $2456623.387(68)$  & $2455520.456(76)$ & $2456623.386(52)$ & $2455761.585(93)$\\
  \rprs 	&  $0.1176\pm0.0013$ & $0.1247\pm0.0022$ & $0.1248\pm0.0018$ & $0.1280\pm0.0030$ & $0.1266\pm0.0013$ & $0.1168\pm0.0022$\\
  $u_1$ 	&  $0.3205$	& $0.2676$ & $0.2676$  & $0.2667$ & $0.2667$ & $0.2508$ \\
  $u_2$ 	&  $0.2884$	& $0.2560$ & $0.2560$  & $0.2595$ & $0.2595$ & $0.2571$ \\
  $\sigma_w$ 	&  $8.4\times10^{-5}$	& $1.0\times10^{-4}$ & $9.1\times10^{-5}$ & $1.0\times10^{-4}$ & $1.0\times10^{-4}$ & $8.2\times10^{-5}$\\
  $\sigma_r$ 	&  $9.0\times10^{-5}$	& $1.2\times10^{-4}$ & $1.8\times10^{-4}$ & $1.8\times10^{-4}$ & $1.1\times10^{-4}$ & $3.4\times10^{-4}$\\
  $\beta$ 	&  $1.47$ &  $1.49$ & $2.25$ & $1.98$ &  $1.47$ & 4.26\\
  \hline
 \end{tabular}
 \linebreak[4]
\linebreak[4]
{\footnotesize $^{\dagger}$ Half transits}
\end{table*}

\subsection{The detection of potassium}
\begin{figure*}
\centering
\includegraphics[width=\textwidth]{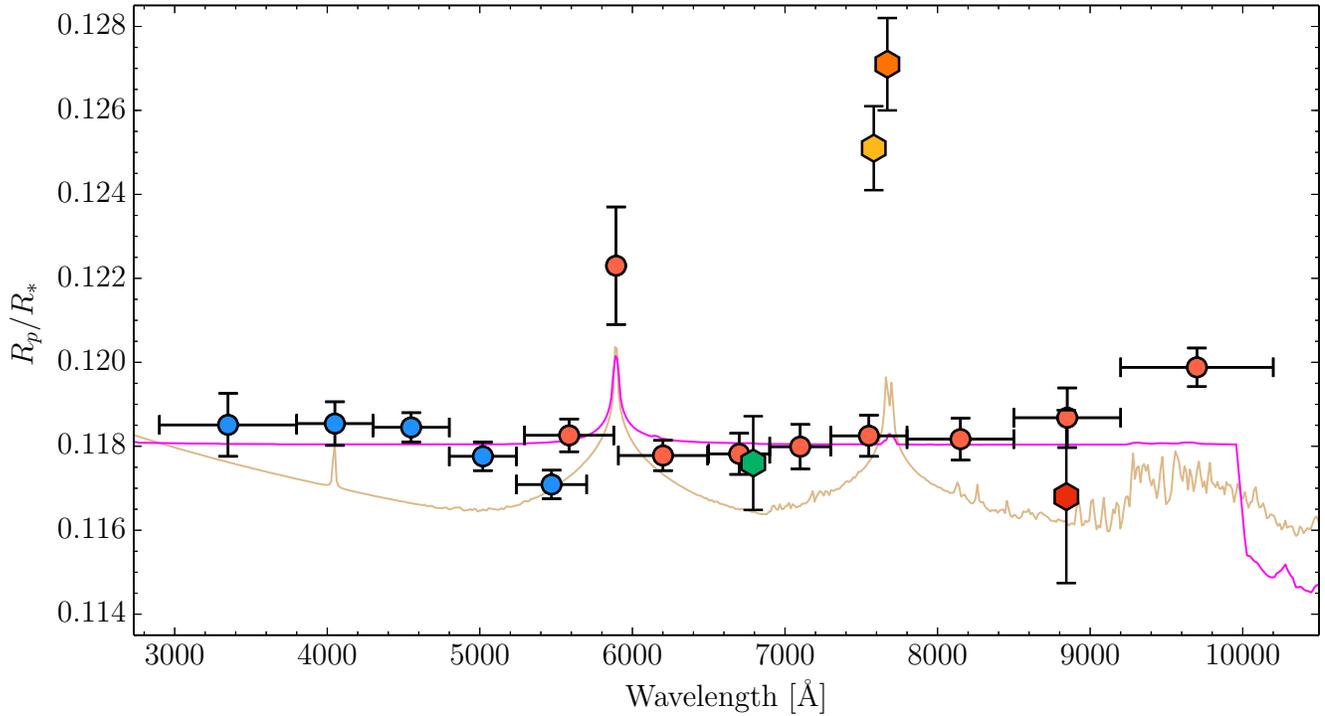}
 \caption[Transmission spectrum of HAT-P-1b.]{Transmission spectrum of HAT-P-1b with data from \cite{nikolov14} shown as blue (STIS/SG430L) and red (STIS/G750L) points with the data from the {\it GTC} shown as hexagonal points. The model transmission spectra are those presented in \cite{nikolov14} and assume an isothermal hydrostatic uniform abundance with a equilibrium temperature of 1200~K. The magenta line is an isothermal model by \cite{burrows10} with an 'extra absorber' at altitude with an opacity of 0.03~cm$^2$ g$^{-1}$ from 0.4 to $1.0~\mu$m. The brown line is a 1200~K isothermal model (without TiO/VO) by \cite{fortney08,fortney10}. Both spectra have been binned using a 12~\AA\ bin.}
  \label{img:hat-p-1b_transpec}
\end{figure*}

We observed four transits, two of which were outside the potassium feature at $6792$~\AA\ ($\mathrm{R}=\lambda / \Delta \lambda =566$) and $8844$~\AA\ (R=737) and two inside the potassium feature probing the line wing at $7582.0$~\AA\ (R=632) and the line peak at $7664.9$~\AA\ (R=639). Both the raw and detrended light curves, together with their best fit models, are shown in Fig.~\ref{img:lcs_6797} and Fig.~\ref{img:lcs} with their best fit model parameters presented in Table~\ref{results}. A transmission spectrum of HAT-P-1b showing all the individual {\it GTC} data points together with the STIS/G750L data from \cite{nikolov14} is shown in Fig.~\ref{img:hat-p-1b_transpec_zoom} and the complete optical transmission spectrum showing a weighted mean of the potassium probing wavelengths at $7582.0$~\AA\ and $7664.9$~\AA\ is shown in Fig.~\ref{img:hat-p-1b_transpec}. Comparing the weighted means of the potassium probing wavelengths to the weighted mean of the observations centred at $6792$~\AA\ and $8844$~\AA, presented in Table~\ref{tabl:results_combined}, the detection of potassium is evident from an increase in the radius ratio of $0.0073\pm0.0017$ at $7582.0$~\AA\  ($4.3\,\sigma$ significance) and $\Delta$\rprs$ = 0.0094\pm0.0016$ at $7664.9$~\AA\ ($6.1\,\sigma$ significance).

The large increase in the radius ratio at the potassium probing wavelengths could be due to an unknown systematic possibly related to the small number of out of transit data points for all the transits. The half transit observations conducted on the 19 of November 2010, displayed a strong potassium detection and it was suspected this could be a systematic related to a correlation between the half transit nature and the large increase in radius ratio observed at the potassium probing wavelengths. This prompted a new set of observations to be done in 2013 at the same wavelengths and using the same setup. The results show a remarkable consistency between the observations done almost three years apart with observations done at 7582~\AA\ showing a difference of \rprs$_{2010-2013} = -0.0001\pm0.0029$ and those done at 7664.9~\AA\ a difference of \rprs$_{2010-2013} = 0.0014\pm0.0033$. The potassium detection is further strengthened when compared to the two {\it GTC} observations outside the feature at 6792~\AA\ (green hexagon in Fig.~\ref{img:hat-p-1b_transpec_zoom}) and 8844~\AA\ (red hexagon in Fig.~\ref{img:hat-p-1b_transpec_zoom}). Even though the 6792~\AA\ observations lack an ingress, and although the the 8844~\AA\ observations suffered from problems with primary mirror distortion causing a clear correlation with FWHM, both observations are highly consistent with the {\it HST} data. The 12~\AA\ bandwidth centered at 7582~\AA\ and 7664.9~\AA\ does not include telluric lines which, if present, may have enhanced the absorption. This is further confirmed by looking at the images themselves where no signs of telluric rings were be seen.

From the {\it HST} observations at low resolution, \cite{nikolov14} detected a sodium line, but did not see either broad wings or a broad potassium feature. Our narrow-band photometric measurements are at higher resolution than the {\it HST} data, with the detected 12~\AA\ potassium feature reaching to higher altitudes than the 30~\AA\ sodium line. For observations at higher altitudes, pressure broadening is expected to become less important, causing a narrow potassium profile. These results echo those of \cite{huitson12} who detected a narrow sodium profile in the atmosphere of HD~189733b only at high resolution, indicating a narrow sodium feature.

\begin{table}
\centering

\caption{Weighted means of the light curve radius ratios for HAT-P-1b.}
\label{tabl:results_combined}
 \begin{tabular}{@{}lc}
  \hline
  \hline
  Wavelength & \rprs\\
  \hline
  Base ($6792$~\AA\ and $8844$~\AA) & $0.1174\pm0.0010$\\
  Wing ($7582.0$~\AA) 	& $0.1248\pm0.0014$\\
  Core ($7664.9$~\AA)	& $0.1268\pm0.0012$\\
  \hline
 \end{tabular}
 \linebreak[4]
\linebreak[4]
\end{table}

\subsection{The effects of stellar variability}
\label{sec:stellar_activity}
Stellar variability is known to alter the flux received from the host star, thereby directly affecting the measured transit depth. Unocculted star spots may increase the transit depths, whereas occulted star spots may cause the transit depth to be underestimated. HAT-P-1 b is not considered to be an active star, showing low chromospheric activity in the Ca\Rmnum{2} H and K lines with $\log(R'_{\mathrm{HK}})=-4.984$ \citep{bakos07,knutson10}. {\it HST}/STIS data show no detectable spot crossings and long term variability monitoring from the ground has resulted in a $0.05$ per cent upper limit on the variability \citep{nikolov14} corresponding to $\Delta$\rprs$\sim0.0005$. Stellar variability is therefore unlikely to be the main cause behind the observed difference between results from data obtained at different epochs. We therefore regard the potassium feature robust and not due to systematics errors which could have been introduced by variability.

\subsection{The effects of resolution}
The {\it HST} observations of \cite{nikolov14} showed no detection of potassium with a 75~\AA\ bin centred at 7682.5~\AA, between the between the D1 and D2 doublets, with a derived radius ratio of \rprs$= 0.11824\pm0.00096$. Compared to the {\it GTC} observations centred near the D2 potassium line (\rprs$ = 0.1268\pm0.0012$ at $7664.9$~\AA), a radius ratio difference of $\Delta$\rprs$ = 0.0086\pm 0.0015$ was observed. Since the {\it GTC} observations near the core of the D2 line at $7664.9$~\AA\ were done at a resolution of $R=\lambda/\Delta\lambda=639$ ($\Delta \lambda \sim 12$~\AA) compared to the {\it HST} observations which were done at a resolution of R=102 ($\Delta \lambda \sim 75$~\AA) in the same wavelength domain, we explored if the effects of resolution can explain these apparently discrepant results.

\begin{figure*}
\includegraphics[width=\textwidth]{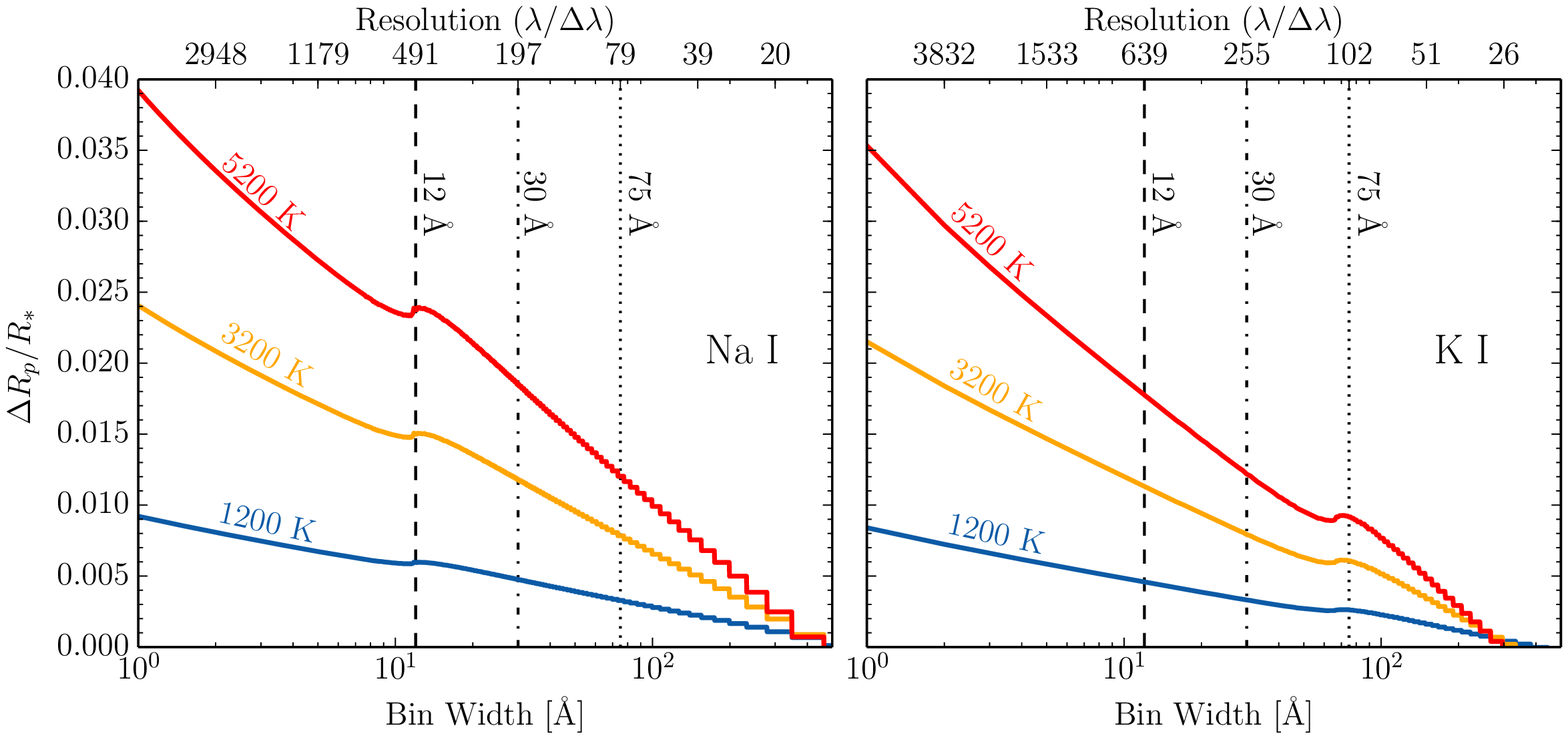}
 \caption[Changes in the planet-to-star radius ratio as function of bin width and resolution.]{Changes in the planet-to-star radius ratio as function of bin width and resolution for a 1200~K (blue), 3200~K (orange) and a 5200~K (red) isothermal model measured by centring a bin on the D2 sodium (left) and a D2 potassium line core (right). The vertical axis are in units of radius ratios above the white light curve radius ratio. The 12, 30 and 75~\AA\ bin widths are indicated by dashed, dot-dashed and dotted lines. The slight increase in radius ratio seen at 12~\AA\ in the left plot and 68~\AA\ in the plot on the right is due to the 5.98~\AA\ and 34~\AA\ separation between the D1 and D2 doublets of the sodium and potassium lines being included in the bin.}
  \label{img:resolution}
\end{figure*}

We calculated the changes in the planet-to-star radius ratio as a function of bin width and resolution for the {\it HST} observations by stepwise binning a high resolution sodium and potassium profile model centred on the D2 sodium and D2 potassium line. The {\it HST} line spread function was modeled using a box function. As the bin width was decreased from 75 to 12~\AA\ a radius ratio increase of $\Delta$\rprs$=0.0020$, $\Delta$\rprs$=0.0053$ and $\Delta$\rprs$=0.0086$ was measured for a 1200~K, 3200~K and 5200~K isothermal potassium line model respectively (see Fig.~\ref{img:resolution}). When compared to the observed radius ratio difference of $\Delta$\rprs$ = 0.0085\pm 0.0012$ between the {\it HST} and {\it GTC} observations, we found that a $\sim5200$~K isothermal model is required to explain why the large potassium absorption was seen in the {\it GTC} data but was not seen in the {\it HST} data. This temperature is however too hot compared to previous temperature measurements, which makes it highly unlikely that resolution alone is responsible for the difference in the planet-to-star radius ratio derived from the {\it GTC} and the {\it HST} data. One possible explanation is that if the potassium is probed higher up in the atmosphere, where the temperature is higher, this would lead to a larger scale height \citep{vidal11, vidal11b}. This explanation is further discussed in \S~\ref{subsec:temperature}.

\subsection{The effects of potassium abundance and mean molecular weight}
The large radius ratio observed in the core of the potassium line at high resolution in the GTC data and not seen in the {\it HST} data at lower resolution, could in part be due to an enhanced abundance of potassium high in the atmosphere. Jupiter and Saturn both show an increase in metals \citep{atreya03,flasar05}, and an enhancement of metals in the atmospheres of hot Jupiter is a well known consequence predicted by the core accretion model \citep[e.g.][]{safronov72,lissauer93,pollack96}. An enhancement in the abundance of potassium high in the atmosphere can however not explain the large absorption measured in the core of the line. This is because an increase in the potassium abundance with altitude would be required with an abundance difference between the core and the base being many orders of magnitudes apart. This is unlikely as the ionisation of potassium is expected to increase with altitude, resulting in a decrease in abundance of neutral potassium at high altitudes. The condensation of potassium in the lower atmosphere, where the temperature is lower, could cause a depletion of potassium at low altitude, thus serving to increase the relative abundance higher up. The most likely condensate to form would be KCl which would happen at a temperature of $\sim 600$ K and at a pressure of $10^{-3}$ bar \citep{morley12} but this is too far away from the conditions of the lower atmosphere to be a viable explanation.

The photon-dissociation by EUV flux breaks molecular hydrogen into atomic hydrogen at pressures $\lesssim10^{-6}$ bars \citep{garcia07,koskinen13} and this increases the scale height, causing an increase in the observed radius ratio. The radius ratio would at most increase by a factor $\sim 2$ if the potassium line observations probed these pressures. Photon-dissociation can therefore only partly account for the large absorption observed.

\subsection{The effects of temperature}
\label{subsec:temperature}
The apparent slope seen towards shorter wavelengths in the {\it HST} data of \cite{nikolov14} could be due to Rayleigh scattering \citep{lecavelier08a,lecavelier2008b}. As the composition of the gases that affect this slope is not well known, the temperature is derived assuming a Rayleigh slope described by a cross section which is inversely proportional to the $4^{\mathrm{th}}$ power of the wavelength. As molecular hydrogen is the most abundant molecule it is used to derive a temperature. \cite{nikolov14} calculate a temperature of $1884\pm603$~K for a pure H$_2$ atmosphere by modeling the Rayleigh slope using the equation of \cite{lecavelier08a}.

Using Spitzer/IRAC secondary eclipse photometry (at 3.6, 4.5, 5.8 and 8.0~$\mu$m) \cite{todorov10} derived an average dayside temperature of $1500\pm100$~K. In the transmission spectrum, \cite{wakeford13} find a $\sim1000$~K isothermal model by \citep{fortney08,fortney10} best fits the water feature detected at 1.45~$\mu$m, whereas \cite{nikolov14} find a 1200~K isothermal model best represents all their optical and near-IR data.

The enhanced absorption measured at the potassium probing wavelengths compared to both the {\it GTC} data and the previous {\it HST} data \citep{nikolov14} could be due to a larger temperature at the high altitude layers where the potassium is being probed. Assuming a temperature of 1200~K at 0.1 bar, a temperature of $3200^{+400}_{-500}$~K at the altitude where the potassium D2 lines are observed would give a change in radius ratio of $\Delta$\rprs$=0.0087$, which is consistent with what is observed. The larger temperature would correspond to the base of the thermosphere that is being probed by potassium at $\sim10^{-3}$ to $\sim10^{-6}$ bars, where a large increase in temperature is not uncommon in solar system planets \citep{lindal81,lindal85}. This transition region between the lower atmosphere and the outermost, hottest layers of the upper atmosphere is referred to as the base of the thermosphere. High upper atmospheric temperatures are particularly relevant to hot-Jupiter planets as predicted by models (e.g., \citealt{yelle04, tian05, garcia07}) and observed in HD~209458b \citep{sing08,vidal11} and HD~189733b \citep{huitson12}. For resolved alkali lines, the temperature can be derived from scale height measurements in both the wings, which probe lower altitudes, and the core, which probes higher altitudes.

\section{Conclusions}
\label{sec:conclusion_hat}
We present four {\it GTC}/OSIRIS transit observations aimed at probing the presence of potassium in the atmosphere of HAT-P-1b. Two separate transit observations, observed three years apart, detect the potassium feature at high confidence providing a detection with a $4.3\,\sigma$ significance at 7582.0~\AA\ and a $6.1\,\sigma$ significance at 7664.9~\AA\, relative to the {\it GTC} measurements outside the feature at 6792~\AA\ and 8844~\AA\ which are both consistent with the {\it HST} data. This is only the second exoplanet (after XO-2b) where the presence of both atmospheric Na and K has been detected. The potassium detection in XO-2b and HAT-P-1b have both been from the ground. We investigate the effects resolution has on the measurements and argue that although the effect is clearly present, it is not large enough by itself to account for the non-detection of potassium in the {\it HST} data at a lower resolution. An increase in potassium abundance with altitude can not explain the strong potassium detection as an increase many orders in magnitude would be required. The strong presence of potassium inferred from the observations could however be influenced by the dissociation of molecular hydrogen into atomic hydrogen thereby decreasing the mean molecular weight and increasing the scale height.

Finally we discuss how an increase in temperature at the altitude where the potassium line is being probed could lead to the strong detection of potassium. It is likely that we detect potassium higher up in the atmosphere where the temperatures are higher and the scale height is larger. Future observations aimed at sampling the slope of the potassium profile, will provide more stringent constraints on the derived temperature at the altitude where the potassium is detected. Such observations will also allow the degeneracies amongst the possible effects responsible for the large absorption in the line core to be minimised.

\section*{Acknowledgments}
We thank the entire GTC staff and in particular Antonio Cabrera Lavers for their help with conducting these observations. This work is based on observations made with the Gran Telescopio Canarias (GTC), installed in the Spanish Observatorio del Roque de los Muchachos of the Instituto de Astrof\'isica de Canarias on the island of La Palma. The GTC is a joint initiative of Spain (led by the Instituto de Astrof\'isica de Canarias), the University of Florida and Mexico, including the Instituto de Astronom\'ia de la Universidad Nacional Aut\'onoma de M\'exico (IA-UNAM) and Instituto Nacional de Astrof\'isica, Optica y Electr\'onica (INAOE). P.A.W, acknowledges support from STFC. The research leading to these results has received funding from the European Research Council under the European Union’s Seventh Framework Programme (FP7/2007-2013) / ERC grant agreement no. 336792. D.K.S, F.P., and N.N. acknowledge support from STFC consolidated grant ST/J0016/1. G.E.B. acknowledges support from STScI through grants HST- GO-12473.01-A. F.P. is grateful for the Halliday fellowship (ST/F011083/1). P.A.W and A.L.E acknowledge the support of the French Agence Nationale de la Recherche (ANR), under program ANR-12-BS05-0012 "Exo-Atmos". The authors would like to thank referee, Ron Gilliland, for his useful comments.

\bibliographystyle{apj}
\setlength{\bibhang}{2.0em}
\setlength\labelwidth{0.0em}
\bibliography{references}
\label{lastpage}
\end{document}